\documentclass[10pt,conference]{IEEEtran}
\IEEEoverridecommandlockouts
\usepackage{cite}
\usepackage{amsmath,amssymb,amsfonts}
\usepackage{algorithmic}
\usepackage{graphicx}
\usepackage{textcomp}
\usepackage{xcolor}
\def\BibTeX{{\rm B\kern-.05em{\sc i\kern-.025em b}\kern-.08em
    T\kern-.1667em\lower.7ex\hbox{E}\kern-.125emX}}

\usepackage{multirow}
\usepackage[indentfirst=false,rightmargin=0pt, leftmargin=0pt]{quoting}

\usepackage{ifthen}

\newboolean{auxinfo}
\setboolean{auxinfo}{true}   

\newboolean{showoutline}
\setboolean{showoutline}{true}

\newboolean{showcomment}

\setboolean{showcomment}{true} 

\ifthenelse{\boolean{auxinfo}}
  {}
  {
    \setboolean{showoutline}{false}
    \setboolean{showcomment}{false}
  }

\usepackage{xcolor}
\usepackage{amssymb}

\ifthenelse{\boolean{showcomment}}
   {\newcommand{\nb}[2]{\fcolorbox{gray}{yellow}{\bfseries\sffamily\scriptsize#1}{\sf\small$\blacktriangleright${\em #2}$\blacktriangleleft$}}
   \newcommand{\working}[1]{\fcolorbox{gray}{yellow}{{\bf #1}\emph{\scriptsize---in progress---}}}
   \newcommand{\TBD}[1]{\fcolorbox{gray}{yellow}{{\bf #1}\textbf{TBD}}} 
  }
  {\newcommand{\nb}[2]{}{}
   \newcommand{\working}[1]{}
   \newcommand{\TBD}[1]{} 
  }

\usepackage{todonotes}
\ifthenelse{\boolean{showoutline}}{
	\newcommand{\outline}[3]{
		~\newline 
		\fcolorbox{red}{white}{
			\parbox{\columnwidth}{
				\ifthenelse{\equal{#1}{}}{
					\ifthenelse{\equal{#2}{}}{
						\noindent\colorbox[rgb]{0.65,0.16,0}{\textcolor[rgb]{1,1,1}{\textbf{Outline}}}
					}{
						\colorbox[rgb]{0.65,0.16,0}{\textcolor[rgb]{1,1,1}{\textbf{Outline -- Responsible: #2}}}
					}
				}{
					\ifthenelse{\equal{#2}{}}{
						\noindent\colorbox[rgb]{0.65,0.16,0}{\textcolor[rgb]{1,1,1}{\textbf{#1 page(s)}}}
					}{
						\colorbox[rgb]{0.65,0.16,0}{\textcolor[rgb]{1,1,1}{\textbf{#1 page(s) -- Responsible: #2}}}
					}
				}
				#3
			}
		}
	}
}{
	\newcommand{\outline}[3]{}
}

\newcommand\defauxcomm[1]{
       \expandafter\newcommand\csname #1\endcsname[1]{\nb{#1}{##1}}
       \expandafter\newcommand\csname WK#1\endcsname{\working{#1}}
       \expandafter\newcommand\csname TBD#1\endcsname{\nb{#1}}
    } 
   
\defauxcomm{KS}
\defauxcomm{MN}

\usepackage[normalem]{ulem}
\ifthenelse{\boolean{showcomment}}
    {\newcommand{\strike}[1]{\textcolor{red}{\sout{#1}}}}
    {\newcommand{\strike}[1]{}}

\begin{document}
\bstctlcite{IEEEexample:BSTcontrol}
\title{Characterizing The Impact of Culture on Agile Methods: The MoCA Model\\}

\author{\IEEEauthorblockN{Michael Neumann}
\IEEEauthorblockA{Hochschule Hannover \\
\textit{Dpt. of Business Information Systems}\\
Hannover, Germany \\
michael.neumann@hs-hannover.de}
\and
\IEEEauthorblockN{Klaus Schmid}
\IEEEauthorblockA{University of Hildesheim \\
\textit{Software Systems Engineering}\\
Hildesheim, Germany \\
schmid@sse.uni-hildesheim.de}
\and
\IEEEauthorblockN{Lars Baumann}
\IEEEauthorblockA{Hochschule Hannover \\
\textit{Dpt. of Business Information Systems}\\
Hannover, Germany \\
lars.baumann@hs-hannover.de}
}

\maketitle

\begin{abstract}
Agile methods are well-known approaches in software development and used in various settings, which may vary wrt.\ organizational size, culture, or industrial sector.
One important facet for the successful use of agile methods is the strong focus on social aspects. We know, that cultural values influence the behaviour of humans. Thus, an in-depth understanding of the influence of cultural aspects on agile methods is necessary  to be able to adapt agile methods to various cultural contexts.
In this paper we focus on an enabler to this problem. We want to better understand the influence of cultural factors on agile practices. 
The core contribution of this paper is MoCA: A model describing the impact of cultural values on agile elements. 
\end{abstract}


\section{Introduction}
The use of agile methods has steadily grown over the past two decades. Today, agile methods are state of the art approaches in software development \cite{VersionOne.2022}. The need to use agile methods is often argued based on being able to react to dynamic market conditions and reducing time-to-market.
Thus, agile methods are used in many different contexts such as different organizational sizes or industry sectors and in several organizational settings like globally distributed  software development teams \cite{Jalali.2010}. 

Agile methods focus on social aspects like communication and collaboration among the team members and stakeholders. The successful use of agile methods in practice rely on the application of their underlying values and principles \cite{Misra.2009}, which are defined in the agile manifesto \cite{Beck.2001} and, in particular, in the guidelines of specific agile methods like Scrum \cite{Schwaber.2020} or XP \cite{Beck.2000}. Thus, one may assume that the cultural influence on human behaviour and values are of high importance while adopting and using agile methods in practice. However, neither in practice nor in research did we find a systematic approach which describes how cultural aspects influence the elements of agile methods like practices or roles in order to provide a basis for a successful application of agile methods in different cultural settings.

The core contribution of this paper is the introduction of an initial causal model, which aims to describe the cultural impact on the elements of agile methods in a systematic manner. The \textbf{M}odel \textbf{o}f \textbf{C}ultural Impact on \textbf{A}gile Methods (MoCA) aims to provide a basis for a better understanding of the need of adapting and using the elements of agile methods in practice. 

MoCA can be applied both for practical and theoretical purposes. For practical applications the model can be used to adapt an agile approach on a project team level. 
From a theoretical perspective, MoCA aims to provide a deeper understanding of cultural impact on agile elements and the relation among the various factors. 

The paper is structured as follows: We present the related work in Section \ref{Sec2:RelWork}. Based on an explanation of the model´s structure in Section~\ref{Sec3:ModelStructure}, we describe the dimensions of the model. 
In Section~\ref{Sec5:MoCA} we introduce the model of cultural impact on agile elements. The paper closes with a conclusion and future work in Section \ref{Sec7:ConclusionAndFutureWork}.

\section{Related Work}
\label{Sec2:RelWork}
We searched for both, primary and secondary studies to find causal models which describe the cultural impact on the elements of agile methods. As the contributions in this field are limited, we focused our literature search to especially identify  empirical studies dealing with the influence of cultural aspects on agile methods.

We  found only one causal model, which describes the impact of agile practices on the characteristics of a software development result: The \textit{agile practices impact model} presented by Diebold and Zehler contributes a basis for the application of an agile capability analysis~\cite{Diebold.2015a}. 
The authors present an approach for characterising which agile practices impact specific characteristics like the quality of a product, development costs and time, or the integration of customers. 
Their proposed model is structured into levels.
Related to these agile capability  levels the authors also describe possible impact factors on agile practices. Their model lacks the cultural aspect and sets the focus on the impact of agile practices on software development outcome. However, their model presented provides an important contribution and supported us for the conceptualization of our model. 


We also identified several studies dealing with various aspects of culture in the field of agile software development (e.g., \cite{Ayed.2017,Smite.2020,Smite.2021}).

Smite et al.\ address the cultural influence on distributed operating DevOps teams in Sweden and India \cite{Smite.2020} in their multiple case study. 
Based on their preliminary findings, they extended their study and present in \cite{Smite.2021} twelve cultural barriers. Six of these barriers were classified as impediments to the agile practices in use with the  five investigated DevOps teams. The authors also provide recommendations aiming to overcome the cultural barriers. 

Furthermore, several authors point to the importance of considering the different levels of culture like regional or organizational (on the department, team or individual level) \cite{Gupta.2019,Siakas.2007}. Most of these studies focus on the effects of the organizational culture on an agile transformation or the use of agile methods in practice~\cite{Iivari.2011,Strode.2009,Tolfo.2011}. 
Interestingly, the majority of these studies were published more than a decade ago and we know, that agile methods have been evolved (see Section~\ref{Sec3-1:Elements of Agile Methods}). Some authors recommend to react to these circumstances by adapting the organizational culture by transforming from a hierarchical culture to more egalitarian one, which is characterized by a greater emphasis on aspects like collaboration and teamwork (e.g., \cite{Gelmis.2022}).

In summary, we found several studies dealing with the challenges related to cultural characteristics while using or transforming to agile methods. However, we did not find any study, which presents a systematic characterization of the relation between culture and agile methods, especially in the form of a causal model. 

\section{Structure of the Model}
\label{Sec3:ModelStructure}
In this section, we explain the structure of MoCA which will be helpful for an understanding of our presentation of the model in the following parts. The structural ground of MoCA are two dimensions and the hypothetical relations between them. Both dimensions were defined based on thorough analysis. We introduce the dimensions in the following subsections.

\subsection{Agile Elements Dimension}
\label{Sec3-1:Elements of Agile Methods}
In order to characterize the cultural impact on ASD, we must precisely define how to describe individual approaches. In the literature, already several characterizations of agile development were given \cite{SweBok.2014}. They vary in terms of levels of detail and their understanding of concepts to characterize. Thus, we will first aim to explain the kind of agile elements, we are interested in, and then we will systematically analyze these elements based on recent literature.

We know that the authors of well-known agile methods like Scrum \cite{Schwaber.2020} or XP \cite{Beck.2000} follow the principles of software process models \cite{SweBok.2014} as their guidelines describe role models, specific activities, and artifacts. Various authors describe specific approaches and frameworks for agile adoption in practice (e.g., \cite{Diebold.2015b,Sidky.2007}). The wide-spread use of agile methods led to the definition of a multitude of elements due to the tailoring of agile methods to specific contexts \cite{Diebold.2014}. 
The need for customizing agile methods to different contexts is described in the literature by several authors (e.g., \cite{Dingsoyr.2014}). As our paper deals with the cultural impact on specific elements of agile methods, it is necessary to formalize these approaches in order to be able to describe the impact of culture independent of individual methods. As a basis for this, we present such a taxonomy, named the tree of agile methods~\cite{Neumann.2021a}. The creation of the taxonomy is based on a literature review and follows the systematic approach by Usman et al.~\cite{Usman.2017}. The taxonomy describes the formal structure of three dimensions of agile elements (roles, artifacts, and agile activities) and three categories for the dimension of agile activities (agile practice, techniques, and tools). 

As our taxonomy provides a systematic and formalized overview of the elements of agile methods, we decided to use it for creating the first dimension of our model. In a next step, we needed an overview of agile practices and other elements of agile methods. Thus, we performed a tertiary study in order to identify overviews of agile practices. Based on the results, we created the integrated list of agile practices~\cite{Neumann.2022}. The list consists of 38 agile practices which are structured in five categories and provide a systematic overview of agile practices. We use the list to define the dimension of agile elements, in particular on the agile practices level. For the other levels (roles and artifacts), we decided in a first step to analyze and use the guidelines of the well-known agile methods Scrum and XP.

\subsection{Cultural Dimension}
\label{Sec3-2:Culture}
We provide an analysis of cultural characteristics to describe cultural influences on agile elements. We first explain how different cultures can be described with the help of models and why different cultural models exist. We chose  national and organizational culture as our focus and  selected the model of cultural dimensions~\cite{Hofstede.2001} and the competing values model~\cite{Quinn.1983} as our reference point. Finally, we describe our approach to using these models within MoCA. 

The term culture is used and defined in numerous different contexts. Thus,  various different definitions and models exist. In particular, culture is described at different levels like regions, organizations, or professions. Several authors point out that the context under study should be taken into account when defining cultural aspects~\cite{Olie.1995,Kroeber.1952}. 

Various models were presented in literature to describe cultures or to make them comparable. Also, these models describe specific cultures such as national, regional or organizational cultures. In accordance with literature, we assume, that certain cultures affect the behaviour of individual team members~\cite{Karahanna.2006}. According to our SLR~\cite{Neumann.2020}, we analyzed, which cultural models are used in the context of software engineering (see Table~\ref{Table1:OverviewCulturalModels}).
Based on our findings, we decided to focus on the cultural levels, which are commonly used in the field of software engineering: National and organizational culture. 

\begin{table}[b] 
\centering
\vspace*{-1em}\begin{tabular}{l|p{5cm}}
\hline
\textbf{Cultural level} & \textbf{Model}\\
\hline
\multirow{3}{*}{National culture}
 & Cultural Dimensions \cite{Hofstede.2001}\\
 & GLOBE Study \cite{House.2002} \\
& Cultural Dimensions \cite{Trompenaars.2012}\\
\hline
\multirow{3}{*}{Organizational culture} 
& Competing Values Model (CVM) \cite{Quinn.1983}\\
& Adapted iceberg model (Schein) \cite{Tolfo.2011} \\
& C.HI.D.DI. Typology of organizational cultures \cite{Siakas.2007} \\
\end{tabular}
\caption{Known cultural models in software engineering \cite{Neumann.2020}}
\label{Table1:OverviewCulturalModels}
\end{table}

All cultural models characterize culture in some way. Hofstede created questionnaires to determine the cultural attitude of respondents~\cite{Hofstede.2013} based on multiple characteristics, to which he refers as cultural values. For each characteristic questionnaires are defined that operationalize the measurement of each characteristic. 
The results of these measurements are given on a scale from 0\ldots 100.  For example, the model by Hofstede consists of six dimensions. Due to the limited space we will focus on three dimensions and provide an explanation based on~\cite{Hofstede.2001} below as the understanding of these dimensions will be helpful for the following parts of the paper. A detailed description of the other dimensions is given in~\cite{Hofstede.2001}.

\textit{Masculinity vs.\ Feminity (MAS):} This dimension describes values based on stereotypes commonly associated with masculinity and femininity.  High values indicate a masculine culture, which means, that achievement of success, power, or performance are highly valued. Feminine societies (low MAS value) emphasize relationships and collaboration. This dimension correlates with the IDV dimension.

\textit{Power Distance Index (PDI):} The power distance represents the extent to which individuals with less power accept or expect an unequal distribution of power. An example of this is the behavior towards superiors in a company. Societies with a high PDI value believe in the hierarchy and take its orders without questioning them. Societies with a low PDI value  strive towards equality in terms of the distribution of power.

\textit{Uncertainty Avoidance (UAI):} This dimension describes whether members of a culture feel comfortable or uncomfortable in new or unknown situations. If there is a high degree of uncertainty avoidance, national culture is characterized by clear regulations (such as laws or security measures) and tries to create structures that are as clear as possible. In national cultures that show a low level of uncertainty avoidance, there is tolerance for other opinions. Besides, regulations are less less precise and strict.

In order to avoid confusion and remain in line with established software engineering terminology, we use the term \textit{metric} to refer to the measurement of these characteristics and the term \textit{value} to refer to a specific value of such a metric. We use the six metrics defined in this model as one part of our dimension and determine their values by using his questionnaires~\cite{Hofstede.2013}. These questionnaires were validated in several studies~\cite{Hofstede.2001,HofstedeMinkov.2013}.

According to the second relevant cultural level, the organizational culture, we analyzed the used cultural models in the field of software engineering related to the cultural dimensions described by Hofstede in order to be able to consider other cultural aspects, which are not covered by Hofstede. Based on this analysis we decided to use the competing values model (CVM) presented by Quinn and Rohrbaugh \cite{Quinn.1983}, because it has been validated \cite{Howard.1998}, used in the field of software engineering and provides new cultural characteristics to our model~\cite{DueholmMuller.2013}. The CVM covers two cultural dimensions: organizational preference for structure (OPS) and organizational focus (OF). 

Thus, our second dimension of the model consists of eight cultural metrics: \textit{Power Distance Index}, \textit{Uncertainty Avoidance}, \textit{Invididualism}, \textit{Masculinity}, \textit{Long Term Orientation}, \textit{Indulgence}, \textit{Organizational preference for structure} and \textit{Organizational focus}.

\section{Cultural Impact Model}
\label{Sec5:MoCA}


\begin{figure}[!tb]
\centering
\includegraphics[scale=0.39]{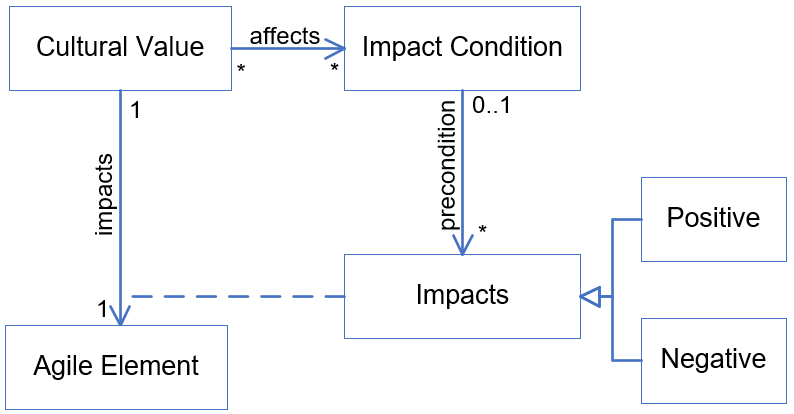}
\caption{MoCA meta-model}\vspace*{-1em}\label{fig1:MoCA-meta}
\end{figure}
In this section, we introduce the theoretical \textbf{M}odel \textbf{o}f \textbf{C}ultural Impact on \textbf{A}gile Elements (MoCA). MoCA consists of a causal model and a meta-model. The MoCA causal model describes the impact of cultural values on the successful use of agile elements. We did also create a meta-model to formalize the structure of the MoCA causal model. First, we will explain our motivation to create MoCA followed by a discussion of the concept of a causal model and the reasons why this modeling is suitable for our purpose. Next, we describe our approach to create MoCA and explain the MoCA meta-model. Finally, we present an example from the MoCA causal model. 

Cultural characteristics can affect the behaviour of people. Important aspects for a successful use of ASD, like communication skills, can be influenced by cultural characteristics. 
Our vision is to  adapt and combine agile elements according to the cultural context for a specific development. As an important first step in this direction we created an initial causal model. This model describes the  impact of cultural characteristics on the successful use of agile elements. 
The model is based on the dimensions of agile elements (see Section~\ref{Sec3-1:Elements of Agile Methods}) and culture (see Section~\ref{Sec3-2:Culture}).

Causal models are mostly used to describe a causal relationship between variables of interest based on a systematic structure~\cite{Russo.2011}. 
This type of modeling has been used in software engineering, e.g., to describe the impact of characteristics of global software development on project goals~\cite{Lamersdorf.2010}. Also, causal models are used in ASD~\cite{Diebold.2015a}.

As a starting point, we created hypotheses about the most likely impact of culture on the various agile elements. We based this on an analysis of existing studies (especially \cite{Ayed.2017,JavdaniGandomani.2016,Smite.2020,Smite.2021}). Further, the first author of the paper has multi-year experience in coaching ASD teams in international contexts. These experiences were also implicitly taken into account during the hypothesis generation. We also analyzed the guidelines of the agile methods Scrum \cite{Schwaber.2020} and XP \cite{Beck.2000}. A systematic analysis of all potential combinations of cultural characteristic and agile elements was performed by creating the full two-dimensional matrix with 384 possible combinations and explicitly considering arguments for a positive and a negative impact\footnote{The two-dimensional matrix is available at Figshare~\cite{Neumann.2023}.}.

The impact of cultural characteristics on the successful use of agile elements can be positive or negative, which means that for a positive influence a high value of a cultural metric (e.g., high PDI) would be positive for the success of the corresponding agile element, while a low  value would be negative. In the case of negative influence, this relationship is inverted.
The impact of multiple values of cultural metrics would be additive (however, this is meant in a more qualitative way; we do not assume that a numerical addition would be possible).
Causal impact can also be conditional. This means that the influence is only assumed to exist, if this condition is true, otherwise the cultural characteristics will not impact the agile element. Preconditions do not need to exist for all causal relationships. In these cases, we omitted the condition (IF) altogether. 

In order to provide a systematic and consistent documentation of cultural impact on agile elements, we used the following template:

\begin{quoting}
   IF (\textit{CONDITION}) THEN \\
   \hspace*{1.5em}\textit{VALUE OF CULTURAL METRIC} $\rightarrow_{\scriptsize\textit{(IMPACT)}}$ \textit{CHARACTERISTIC}
\end{quoting}

The MoCA meta-model provides an abstract description and  formalizes the structure of the MoCA causal model (see Figure~\ref{fig1:MoCA-meta}). 
Inspired by the model from Diebold and Zehler~\cite{Diebold.2015a}, we use the Unified Modeling Language (UML) to specify and visualize it in Figure~\ref{fig1:MoCA-meta}. 
The  impact of a \textit{Cultural Value} on an \textit{Agile Element} is represented by the association \textit{impacts} and is further refined by the corresponding association class, which has \textit{positive} and \textit{negative} subclasses. Preconditions for these impact relationships are represented by the class \textit{Impact Condition}.
The impact condition can reference multiple cultural values and is related as a precondition to the \textit{Impacts}-class. For example, a cultural value of \textit{high power distance} impacts the agile element \textit{daily meeting}, if the impact condition \textit{manager  attends meeting} exists.

\begin{figure}[!tb]
\centering
\includegraphics[scale=0.8]{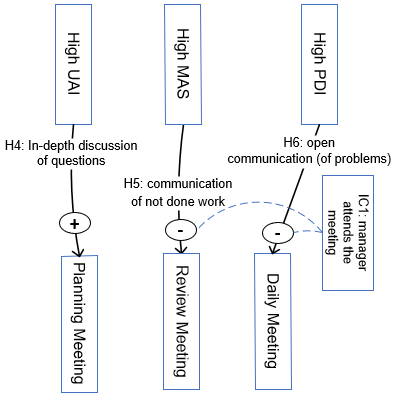}
\caption{Example of the MoCA}\vspace*{-1em}\label{fig2:MoCA-example}
\end{figure}

Because of its size, a full visualization of the MoCA causal model is not possible. However, we show a graphical excerpt of the MoCA (see Figure~\ref{fig2:MoCA-example}). The example presents three cultural values at the top, three agile elements at the bottom and the causal impact between them. The specific causal impact is visualized by an arrow between them, the impact is refined by the sign in the circle (plus for positive, minus for negative impact). The example also covers one impact condition (on the right side) for two impact relations. Below, we describe three relations on detail.

\textit{H4 in-depth discussions of questions:} A high UAI value refers to a pronounced uncertainty avoidance, which leads to a positive impact on a planning meeting, because the team members are avoid of discussing open questions in order to get a deeper understanding of the product requirements. No precondition is needed.

\textit{H5 communication of done work:} A high MAS value represents cultural characteristics like the focus of success and power. Thus, individual failures like (partially) not done work, may not be communicated, when a manager is attending the meeting. According to this characteristics, we assume a negative impact on the agile element review meeting. The impact condition \textit{IC1: manager attends the meeting} is needed for this hypothetical impact.

\textit{H6 open communication (of problems):} If the impact precondition manager attends the meeting is given, a high value of power distance index leads to a negative impact on the daily meeting because the team members may not be willing to communicate problems. 

It is important to mention, that the causal relationships in the conceptualization of the MoCA are hypothetical but follow a systematic approach. In a first step, we defined and filled the above described two dimensions. Secondly, we described the causal relationship between the two dimensions based on findings from the literature and our own experience in the field of agile software development. Finally, we created a meta model of the MoCA in order to be able to describe the formal structure of the model.

\section{Conclusion and Future Work}
\label{Sec7:ConclusionAndFutureWork}
In this paper, we presented the Model of Cultural Impact on Agile Elements (MoCA), describe our motivation for its creation, the conceptualization of the model and how to apply it. 


MoCA consist of two dimensions: Culture and Agile Elements. We used a systematic approach to create both dimensions based on extracted findings from the literature. The dimension of agile elements consists of 38 agile practices and 10 roles. The cultural dimension was created based on a thorough analysis of well-known comparative and descriptive cultural models, which are used in the area of software engineering. The dimension consists of eight cultural metrics. The hypothetical relation between both dimensions is represented by an impact, which we designed based on the findings from the literature and our own experience. In total, MoCA covers 384 possible combinations in its two-dimensional matrix and the arguments for a positive or negative impact. 

Although we followed a systematic approach for the conceptualization of the MoCA, we see the need for its validation based on empirical findings to provide evidence on the specific relations. In order to be able to validate the MoCA, we already prepared and conducted case studies in Japan and Germany. Currently, we are analyzing the results and aiming to present them in the near future. 



\bibliographystyle{IEEEtran}
\bibliography{references}

\end{document}